\documentclass{iopconfser}
\usepackage{lmodern}
\usepackage{graphicx,hyperref}
\usepackage{amsmath, amssymb, amsfonts}
\usepackage{xcolor}
\usepackage{ulem}

\usepackage{algorithm}
\usepackage{algpseudocode}
\newcommand{\D}{\mathrm{d}}

\begin{document}

\title{Efficient Simulation of Electron-Positron Pair Production in Foam Targets in the low $\chi$-Regime}

\author{Oliver Mathiak$^{1}$, Lars Reichwein$^{2,1}$ and Alexander Pukhov$^{1}$}
\affil{$^1$Institut für Theoretische Physik I, Heinrich-Heine-Universität Düsseldorf, 40225 Düsseldorf, Germany}
\affil{$^2$Peter Grünberg Institut (PGI-6), Forschungszentrum Jülich, 52425 Jülich, Germany}

\email{oliver.mathiak@hhu.de}

\begin{abstract}
The generation of electron-positron pairs using direct laser-accelerated electrons and a cone-shaped reflector target for the generation of strong electromagnetic fields is investigated using particle-in-cell simulations. A newly implemented sub-sampling routine for the code \textsc{vlpl} is presented which allows for a better description of quantum electrodynamical processes which would otherwise come at a high computational cost.
\end{abstract}

\section{Introduction}
With the advent of petawatt-class laser systems, such as ELI \cite{Mourou2011}, XCELS \cite{Khazanov2023} and SEL/SULF \cite{Liang2020}, probing new frontiers of ultra-high-energy laser-plasma interactions has come into reach. With such intensities, it is possible to approach the critical Sauter-Schwinger field $E_\mathrm{cr}=mc^3/e\hbar$ corresponding to an intensity $\sim 10^{23}$ W/cm$^2$. In these regimes, strong field quantum electrical (SF-QED) effects not only become relevant, but might even dominate the dynamics of the system. Thus, classical descriptions of the interaction of particles with the fields are no longer sufficient, but a quantum-mechanical description is necessary.
Such SF-QED effects are commonly taken into account in particle-in-cell (PIC) codes by means of a semi-classical Monte Carlo scheme \cite{Elkina2011}.

As of now, various schemes for generating abundant electron-positron pairs have been proposed. Filipovic \textit{et al.} proposed a setup in which two counter-propagating laser pulses irradiate a solid target at grazing incidence to extract a large number of electrons which can produce a QED cascade when colliding with the counter-propagating laser \cite{Filipovic2022}.  Samsonov \textit{et al.} studied the generation of strong fields via the Inverse Faraday Effect when a circularly polarized laser pulses irradiates a solid cone target \cite{Samsonov2025}. The mechanism was shown to generate a long-lasting electron-positron plasma.

\begin{figure}[h!]
    \centering
    \includegraphics[width=\textwidth]{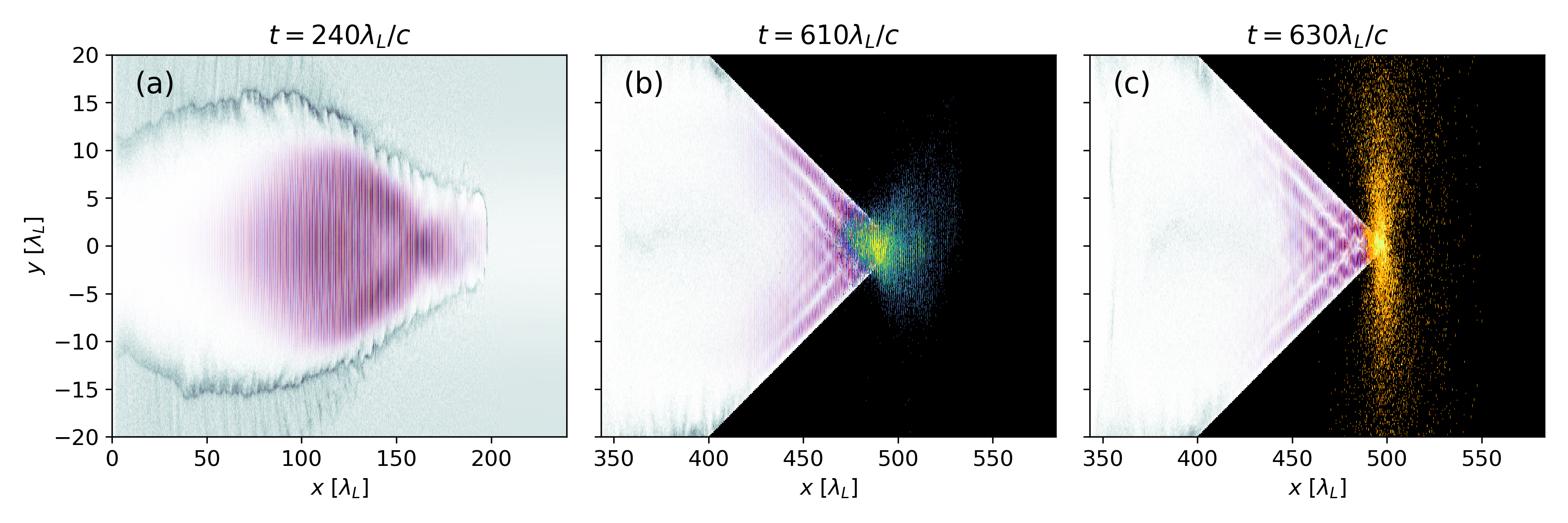}
    \caption{Depiction of the setup in PIC simulations. (a) The laser (red/blue) propagates through the homogenized foam (grey), accelerating electrons via DLA. (b) The laser pulse is reflected at the cone target (black) and high-energy photons (green) are produced. (c) The photons subsequently decay into electron-positron pairs (orange).}
    \label{fig:scheme2d}
\end{figure}

In this proceeding, we investigate the generation of electron-positron pairs using DLA in foam targets using PIC simulations. The driving laser pulse propagates through a foam, accelerating abundant electrons via DLA (cf. Fig. \ref{fig:scheme2d}). Subsequently, the laser reaches a solid, cone-shaped target which reflects the pulse. The electrons produce high-energy photons via non-linear inverse Compton scattering which can decay into electron-positron pairs in the non-linear Breit-Wheeler process.  In the following section, we describe the SF-QED relevant in our parameter regime and discuss their numerical implementation in section \ref{sec:sfqed}. In particular, we present a subsampling algorithm that avoids the high computational cost of generating a high number of photons that do not subsequently decay into electron-positron pairs (and therefore would be considered not of interest). A comparison of subsampled and non-subsampled PIC simulation results for our proposed target geometry is presented in section \ref{sec:results}.

\section{SF-QED Effects and Sub-Sampling Routine}\label{sec:sfqed}

In the following, we first describe the SF-QED effects relevant to our parameter regime. Subsequently, we detail the implementation of the sub-sampling routine in the particle-in-cell (PIC) code \textsc{vlpl} \cite{Pukhov1999}.

\subsection{Strong-Field Quantum Electrodynamics}
The SF-QED effects arising at high intensities are the emission of a high-energy photon by an energetic lepton and the subsequent decay of such a photon into an electron-positron pair.
The strength of QED processes is determined by the dimensionless quantum non-linearity parameter
\begin{align}
    \chi = \frac{e\hbar}{m_e^3c^4} \sqrt{(F_{\mu\nu} p^\nu)^2} 
    = \frac{\gamma}{E_\mathrm{cr}}\sqrt{(\mathbf{E} + \mathbf{v} \times \mathbf{B})^2 - (\mathbf{E} \cdot \mathbf{v}/c)^2} \; ,
\end{align}
where $e$ is the electron charge, $\hbar$ the reduced Planck constant, $m_e$ the electron rest mass and $c$ the speed of light. Further, let $F^{\mu\nu}$ be the electromagnetic field tensor, $p^\nu$ the particle's four-momentum and $\gamma$ the corresponding Lorentz factor.

In the locally constant crossed field (LCFA) approximation, where we assume the fields to be orthogonal $\mathbf{E} \perp \mathbf{B}$ and locally constant in time $\mathbf{B} (t) = \mathbf{B}_t$ over some time $\delta t$, the effects can be approximated as a single-particle interaction with the background field.
The rate of photon emission via non-linear inverse Compton scattering can thus be written as \cite{Ritus1985-sq}
\begin{align}
    \frac{\D W_\gamma}{\D \chi_\gamma} = \frac{2}{3} \frac{\alpha^2}{\tau_e} \frac{G(\chi, \chi_\gamma)}{\chi_\gamma}
\end{align}
with the fine structure constant $\alpha = e^2 /4\pi\epsilon_0\hbar c$ and the time associated with the classical electron radius $\tau_e = r_e/c$, with $r_e = e^2/4\pi\epsilon_0m_ec^2 \approx 2.82$ fm. Here, $\chi$ and $\chi_\gamma$ are the quantum nonlinearity parameters of the emitting lepton and the photon, respectively.
The so-called quantum emissivity (or ``Gaunt factor'') is defined as 
\begin{align}
    G (\chi, \chi_\gamma) = \frac{\sqrt{3}}{2\pi} \frac{\chi_\gamma}{\chi} \left[ \int_\nu^\infty K_{5/3} (y) \; \D y + \frac{2}{3}\chi_\gamma \nu K_{2/3}(\nu)\right]
\end{align}
with $\nu = 2\chi_\gamma / [3\chi(\chi - \chi_\gamma)]$ and $K_\nu$ being the modified Bessel function of the second kind.
This can be approximated as \cite{10.1007/3-540-55250-2_37}
\begin{align}
    W_\gamma = \frac{5}{2\sqrt{3}} \frac{\alpha \chi_\gamma}{r_e \gamma} \frac{1}{\sqrt{1 + \chi_\gamma^{2/3}}} \propto \chi_\gamma^{1/3} \; .
\end{align}
A high-energy photon can interact with multiple background photons to decay into an electron-positron pair. The corresponding rate is given by \cite{Ritus1985-sq}
\begin{align} \label{eq:BWRate}
    \frac{\D W_\mathrm{PP}}{\D \varepsilon_\pm} = \frac{\alpha m_e^2 c^4}{\sqrt{3}\pi \hbar \varepsilon_\gamma} \left[
        \int_\xi^\infty K_{1/3}(y) \;  \D y + \frac{\varepsilon_+^2 + \varepsilon_-^2}{\varepsilon_+ \varepsilon_-} K_{2/3}(\nu)
    \right] \; ,
\end{align}
where $\varepsilon_\gamma$ is the energy of the initial hard photon and $\varepsilon_\pm$ is the energy of the created positron and electron respectively, with $\varepsilon_+ + \varepsilon_- = \varepsilon_\gamma$.
Similarly, one can find an approximation of the total pair creation rate \cite{10.1007/3-540-55250-2_37},
\begin{align} \label{eq:PPApprox}
    W_\mathrm{PP} \approx \alpha m_e^2/ \omega \begin{cases}
        0.23 \chi_\gamma \exp(-8/3\chi_\gamma) & \chi_\gamma \ll 1 \\
        0.38 \chi_\gamma^{2/3} & \chi_\gamma \gg 1
    \end{cases} \; .
\end{align}
Evidently, the pair production rate is exponentially suppressed for $\chi \ll 1$.
This is typically implemented in a classical PIC scheme as a quasi-classical Monte Carlo routine of a single-particle process. The rate \eqref{eq:BWRate} is used as a transition probability between two states in a single simulation time step as $P = R  \D t$.

\subsection{Sub-Sampling Routine for QED Processes}\label{sec:numerics}
\begin{figure}[h!]
    \centering
    \includegraphics[width=0.5\textwidth]{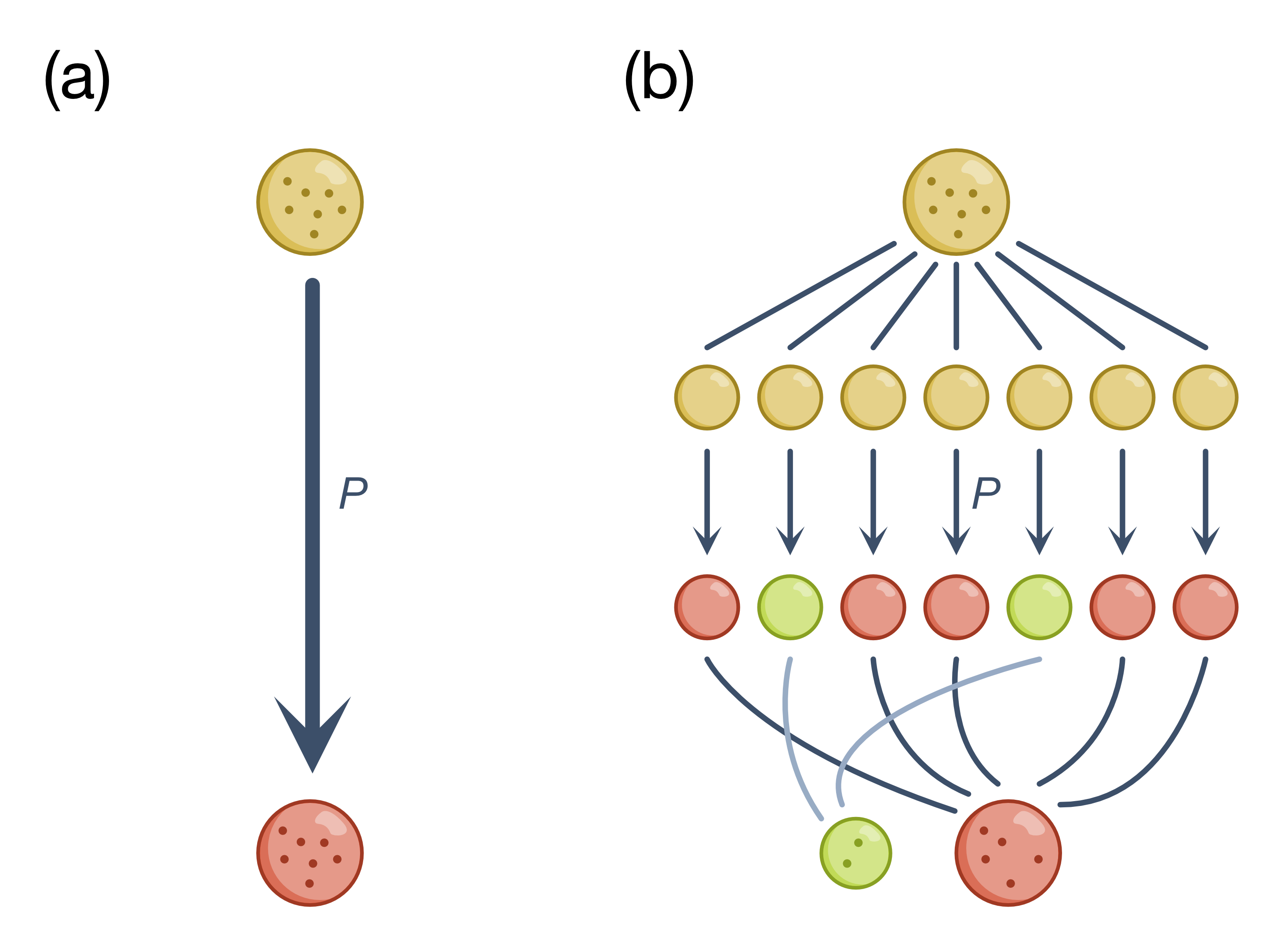}
    \caption{Schematics for the implementation of a process $P$ (a) without sub-sampling and with sub-sampling (b).}
    \label{fig:subsampling}
\end{figure}

From \eqref{eq:PPApprox} it is evident that for small $\chi \ll 1$ the pair creation rate is much more strongly suppressed than the photon emission rate. In practice, that means that a lot more photons get created than pairs. Because the photons barely interact with the background plasma, their overall behaviour is largely uninteresting for PIC simulations, but still increases the computational cost.
Therefore, we have implemented a subsampling routine which improves the computational efficiency considerably by decoupling the granularity of the photon and lepton macroparticles. The scheme is depicted in Fig. \ref{fig:subsampling}. 

Because a macroparticle generally represents many physical particles, it is possible to split a macroparticle into an arbitrary number of sub-particles. Each of these sub-particles may be simulated individually with regard to the SF-QED effects, without loss of physicality.
Within a timestep, we can split a macroparticle into $N$ sub-particles, do the random sample for each sub-particle individually and sum up the resulting photons and leptons. This leaves us with two macro-particles, each with a lower weight (i.e. number of represented physical particles) than the initial macroparticle, improving the resolution of the Breit-Wheeler process. Because the energy and fields within particles of the same macroparticles are identical, the transition probability is identical as well. This makes the subsampling in practice a number of i.i.d. Bernoulli experiments, which are collectively described by a binomial distribution. Accordingly, we can simplify the subsampling calculation as a single random sample of a binomial distribution. In fact, for $N\to \infty$ this converges to the expectation value $P$. However, if one were to forgo the sampling step altogether and use the expectation value, this would produce a prohibitively large number of macroparticles with little weight, as each macroparticle would produce a new macroparticle in each timestep. Instead, a weight threshold would be necessary under which all resulting macroparticles are neglected. This, however, would distort the statistics of the process. In the case of finite subparticles, this threshold is satisfied naturally.
The implementation of the subsampling algorithm in terms of pseudocode is shown in Alg. \ref{alg:cap}.

\begin{algorithm}[h!]
\caption{Sub-sampling algorithm in the case of Breit-Wheeler pair production.}\label{alg:cap}
\begin{algorithmic}
\Require $N, w, \chi, \varepsilon_\gamma$ \Comment{Split macroparticle into $N$ sub-particles of weight $w$}
\State $P = \text{GetRate}(\chi, \varepsilon_\gamma) \Delta t$

\If{$N > 1$}
    \State $k \in B(N, P)$ \Comment{Sample from a binomial distribution}
    \If{$k = 0$} 
        \State return
    \EndIf
    \State $\text{ratio} = k / N$
    \State $p,\, e = \text{CreatePair}()$
    \State $p.w, \, e.w = \text{ratio} \times w$
    \State $w = (1-\text{ratio}) \times w$
\Else
    \State do normal Breit-Wheeler 
\EndIf
\end{algorithmic}
\end{algorithm}

\section{Simulation results} \label{sec:results}

We conduct 2D and 3D PIC simulations using the code \textsc{vlpl} \cite{Pukhov1999, Pukhov2016} with the newly implemented subsampling routine. The considered setup is the following: a pre-ionized near-critical plasma is placed in front of a solid reflector target with a cone-shaped cavity. We irradiate the plasma with a strong 1.5kJ laser pulse of length 150 fs, spot size 2.5 µm and wavelength $\lambda_L = 800$ nm, corresponding to a normalised laser vector potential $a_0 \sim 194$. The laser pulse both accelerates electrons in the plasma and reflects off the target. It acts as both the acceleration mechanism to produce high-energy electrons and as the strong background field after reflection, which enables the SF-QED effects:
electrons are accelerated to multiple GeV via direct laser acceleration (DLA) and interact with the reflected laser pulse to generate high-energy photons that quickly decay into electron-positron pairs. The scheme is depicted in Fig. \ref{fig:scheme2d}.

\begin{figure}[h!]
    \centering
    \includegraphics[width=0.5\linewidth]{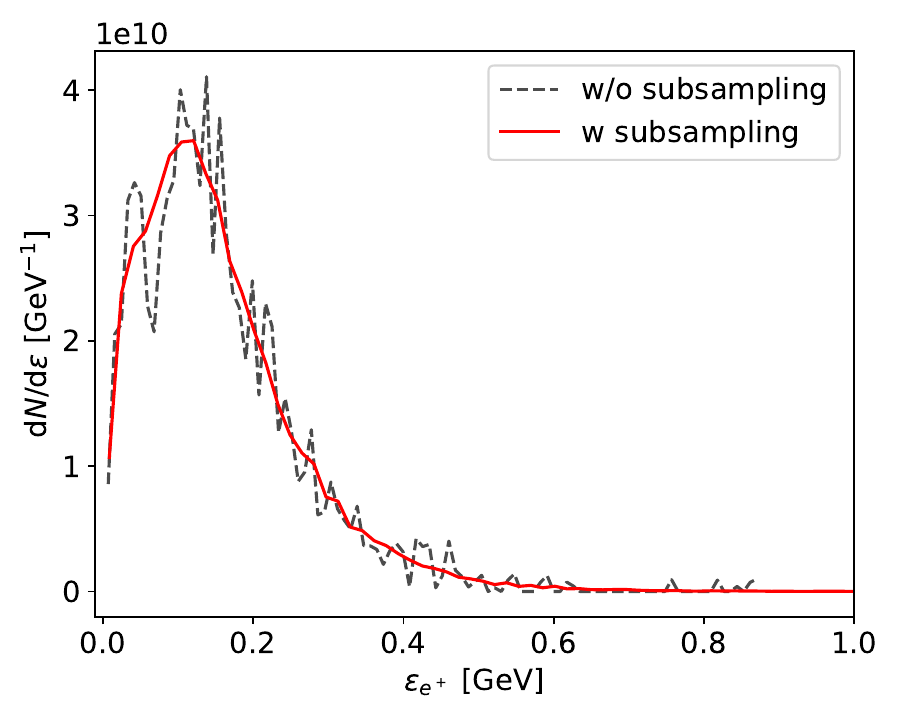}
    \caption{Energy spectrum of the produced Breit-Wheeler pairs with and without subsampling.}
    \label{fig:spectrum}
\end{figure}

We consider a foam plasma of length $L = 400$ µm with a Gaussian channel to guide the laser pulse along the optical axis, retaining its shape and enabling higher field amplification in the cone target. The inner density is $n_0 = 0.3 n_c$ and outer density $n_1 = 1 n_c$ and width $\sigma = 8$µm.
The simulation domain has a size of $240 \lambda_L \times 70 \lambda_L$, with grid resolution $h_x = 0.1 \lambda_L, h_y = 0.15 \lambda_L$. We simulate $4$ macroparticles per cell, such that each macroparticle represents $w \sim 2 \times 10^6$ physical particles. Within our subsampling scheme, we set a minimum weight of $w_\mathrm{min} = 250$, resulting in a split of $N \sim  8000$.
We find that a setup optimized with respect to parameters like channel density and cone angle produces approx. $10^{11}$ electron-positron pairs using a 1.5 kJ laser pulse. A more extensive discussion of the physical results can be found in our separate work \cite{Mathiak2025}.
The resulting positron energy spectrum is shown in Fig. \ref{fig:spectrum}. The subsampling scheme clearly improves the statistics while retaining the overall positron numbers and energy. With our subsampling routine we were able to improve the resolution of the energy spectrum from $\sim1500$ macroparticles without subsampling to $\sim 10^6$ with subsampling.

\section{Conclusions}

In this proceeding, we have studied the effects of strong-field quantum electrodynamics in laser-plasma interactions. We studied the generation of electron-positron pairs in a single-laser setup, using particle-in-cell simulations. Our simple subsampling scheme is shown to vastly improve the resolution of the resulting spectrum while retaining physicality, at a very low computational cost.

\section*{Acknowledgments}
The authors would like to thank Ke Jiang (SZTU) for fruitful discussions.
This work has been supported by the Deutsche Forschungsgemeinschaft and by BMBF (Project 05P24PF1). 
The authors gratefully acknowledge the Gauss Centre for Supercomputing e.V. \cite{GCS} for funding this project
(lpqed) by providing computing time through the John von Neumann Institute for Computing (NIC) on the GCS Supercomputer JUWELS at Jülich Supercomputing Centre (JSC).

\bibliographystyle{ieeetr}
\bibliography{ref_proceeding.bib}

@misc{GCS,
    title = {{Gauss Centre for Supercomputing}},
    note  = {\url{https://www.gauss-centre.eu}}
}

@article{Elkina2011,
  title = {{QED} cascades induced by circularly polarized laser fields},
  author = {Elkina, N. V. and Fedotov, A. M. and Kostyukov, I. Yu. and Legkov, M. V. and Narozhny, N. B. and Nerush, E. N. and Ruhl, H.},
  journal = {Phys. Rev. ST Accel. Beams},
  volume = {14},
  issue = {5},
  pages = {054401},
  numpages = {12},
  year = {2011},
  month = {May},
  publisher = {American Physical Society},
  doi = {10.1103/PhysRevSTAB.14.054401},
  url = {https://link.aps.org/doi/10.1103/PhysRevSTAB.14.054401}
}

@article{Khazanov2023, title={{eXawatt} Center for Extreme Light Studies}, volume={11}, DOI={10.1017/hpl.2023.69}, journal={High Power Laser Science and Engineering}, author={Khazanov, Efim and Shaykin, Andrey and Kostyukov, Igor and Ginzburg, Vladislav and Mukhin, Ivan and Yakovlev, Ivan and Soloviev, Alexander and Kuznetsov, Ivan and Mironov, Sergey and Korzhimanov, Artem and et al.}, year={2023}, pages={e78}}

@article{Mourou2011,
author = {G\'{e}rard Mourou and Toshiki Tajima},
journal = {Opt. Photon. News},
number = {7},
pages = {47--51},
publisher = {Optica Publishing Group},
title = {The Extreme Light Infrastructure: Optics' Next Horizon},
volume = {22},
month = {Jul},
year = {2011},
url = {https://www.optica-opn.org/abstract.cfm?URI=opn-22-7-47},
doi = {10.1364/OPN.22.7.000047},
}

@inproceedings{Liang2020,
author = {Xiaoyan Liang and Yuxin Leng and Ruxin Li and Zhizhan Xu},
booktitle = {OSA High-brightness Sources and Light-driven Interactions Congress 2020 (EUVXRAY, HILAS, MICS)},
journal = {OSA High-brightness Sources and Light-driven Interactions Congress 2020 (EUVXRAY, HILAS, MICS)},
pages = {HTh2B.2},
publisher = {Optica Publishing Group},
title = {Recent Progress on the Shanghai Superintense Ultrafast Laser Facility ({SULF}) at {SIOM}},
year = {2020},
url = {https://opg.optica.org/abstract.cfm?URI=HILAS-2020-HTh2B.2},
doi = {10.1364/HILAS.2020.HTh2B.2},
}

@misc{Mathiak2025,
      title={Electron-positron pair generation using a single kJ-class laser pulse in a foam-reflector setup}, 
      author={Oliver Mathiak and Lars Reichwein and Alexander Pukhov},
      year={2025},
      eprint={2509.15853},
      archivePrefix={arXiv},
      primaryClass={physics.plasm-ph},
      url={https://arxiv.org/abs/2509.15853},
      doi={10.48550/arXiv.2509.15853}
}

@article{Filipovic2022, title={{QED} effects at grazing incidence on solid-state targets}, volume={76}, DOI={10.1140/epjd/s10053-022-00494-4}, number={10}, journal={The European Physical Journal D}, author={Filipovic, Marko and Pukhov, Alexander}, year={2022}, month={Oct}}

@article{Samsonov2025,
    author = {Samsonov, Alexander and Pukhov, Alexander},
    title = {Production and magnetic self-confinement of e-e+ plasma by an extremely intense laser pulse incident on a structured solid target},
    journal = {Matter and Radiation at Extremes},
    volume = {10},
    number = {5},
    pages = {057202},
    year = {2025},
    month = {08},
    issn = {2468-2047},
    doi = {10.1063/5.0260941},
    url = {https://doi.org/10.1063/5.0260941}
}

@article{Pukhov1999, 
    title={Three-dimensional electromagnetic relativistic particle-in-cell code {VLPL} (Virtual Laser Plasma Lab)}, 
    volume={61}, 
    DOI={10.1017/S0022377899007515}, 
    number={3}, 
    journal={Journal of Plasma Physics}, 
    author={Pukhov, A.}, 
    year={1999}, 
    pages={425–433}
}

@Article{Pukhov2016,
  author    = {Pukhov, A.},
  journal   = {CERN Yellow Reports},
  title     = {Particle-In-Cell Codes for Plasma-based Particle Acceleration},
  year      = {2016},
  pages     = {Vol 1 (2016): Proceedings of the 2014 CAS--CERN Accelerator School: Plasma Wake Acceleration},
  copyright = {This work is licensed under a Creative Commons Attribution 4.0 International License.},
  doi       = {10.5170/CERN-2016-001.181},
  publisher = {CERN, Geneva},
}

@ARTICLE{Ritus1985-sq,
  title     = "Quantum effects of the interaction of elementary particles with
               an intense electromagnetic field",
  author    = "Ritus, V I",
  journal   = "J Russ Laser Res",
  publisher = "Springer Science and Business Media LLC",
  volume    =  6,
  number    =  5,
  pages     = "497--617",
  year      =  1985,
}

@InProceedings{10.1007/3-540-55250-2_37,
    author="Yokoya, Kaoru
    and Chen, Pisin",
    editor="Dienes, M.
    and Month, M.
    and Turner, S.",
    title="Beam-beam phenomena in linear colliders",
    booktitle="Frontiers of Particle Beams: Intensity Limitations",
    year="1992",
    publisher="Springer Berlin Heidelberg",
    address="Berlin, Heidelberg",
    pages="415--445",
    isbn="978-3-540-46797-7"
}

\end{document}